\documentclass[12pt]{article}
\usepackage{epsf}
\usepackage{amsmath}
\usepackage{graphics}
\usepackage{cite}

\renewcommand{\thefootnote}{\#\arabic{footnote}}
\begin{document}

\newcommand{\gtrsim}{ \mathop{}_{\textstyle \sim}^{\textstyle >} }
\newcommand{\lesssim}{ \mathop{}_{\textstyle \sim}^{\textstyle <} }

\newcommand{\rem}[1]{{\bf #1}}

\renewcommand{\thefootnote}{\fnsymbol{footnote}}
\setcounter{footnote}{0}
\begin{titlepage}

\def\thefootnote{\fnsymbol{footnote}}

\hfill July 2016\\
\vskip .5in
\bigskip
\bigskip

\begin{center}
{\Large \bf X-events and their Interpretation}

\vskip .45in

{\bf Claudio Corian\`o$^{(a)}$\footnote{email:claudio.coriano@le.infin.it}
and Paul H. Frampton$^{(b)}$\footnote{email: paul.h.frampton@gmail.com\\
homepage:www.paulframpton.org}}

{(a) Dipartimento di Matematica e Fisica "Ennio De Giorgi"\\
 Universit\`a del Salento and INFN Lecce, 
 Via Arnesano 73100 Lecce,Italy}

{(b) 15 Summerheights, 29 Water Eaton Road, Oxford OX2 7PG, UK.}

\end{center}

\vskip .4in
\begin{abstract}
\noindent
We point out that when doubly-charged bileptons are pair produced
at the LHC, kinematics dictate that they are both almost at rest in the
lab frame and therefore their decays lead to final state muons in
a characteristic X-shape with only very tiny track 
curvature because of the high muon energies. Such 
X-events have essentially no standard model
background and provide a smoking gun for the 331 model.  
\end{abstract}

\end{titlepage}

\renewcommand{\thepage}{\arabic{page}}
\setcounter{page}{1}
\renewcommand{\thefootnote}{\#\arabic{footnote}}

\newpage

\section{Introduction}

\bigskip

\noindent
In the aftermath of the discovery of the BEH scalar boson
with $M_H \simeq 125$ GeV, there is an urgent phenomenological
question of what further can be expected to be discovered at the
LHC?

\bigskip

\noindent
The cases for both weak-scale supersymmetry and 
large extra dimensions have been seriously weakened by the
LHC data already in hand, because there is no sign 
of either superpartners or Kaluza-Klein modes.

\bigskip

\noindent
Nevertheless, nobody expects
the Standard Model (SM) to be the final theory. The SM
possesses 28 free parameters, only 6 of which has been
convincingly derived by theory. The 12 fermion masses
remain unexplained.  Leaving temporarily aside the
4 CP-violating phases, the six real mixing angles
in the CKM and PMNS matrices do now have a semi-quantitative
theoretical explanation by appending to the SM a discrete 
nonabelian flavor symmetry. The final six SM parameters
are all mysterious. 

\bigskip

\noindent
In the present article we adopt the strategy
of ignoring the problem of the twenty-two unexplained SM
parameters. Instead we consider the simplest extension
of the SM which has the benefit of explaining why there
are three sequential quark-lepton generations.

\bigskip

\noindent
This is the 331 model invented in 1992 \cite{PHF,PP}
which now appears more unique than expected\cite{CM,CD}. Aside
from trivial redefinitions of the embedding of the electric charge
there is only one known example of such a set-up which can explain
the three-fold replication of the first generation. The model
has far fewer free parameters than occur for either supersymmetry
or extra spatial dimensions. This renders the LHC predictions
more quantitative.

\bigskip

\noindent
An overriding feature of the 331 model is that the symmetry
breaking from 331 to the 321 of the SM must be at a scale
below about 4 TeV. This was first pointed out uniquely in
\cite{PHF}. The theoretical reason involves the group
embedding of the SM $SU(2)_L$ into the 331's $SU(3)_L$ and
arises phenomenologically because of the proximity
of the electroweak mixing angle $\Theta_W$ to the
crucial value above which the embedding of $SU(2)_L$
into $SU(3)_L$ requires an unphysical imaginary
gauge coupling constant.

\bigskip

\noindent
More specifically, at the Z-pole the latest value of $\Theta_W$ is
\cite{PDG}
\begin{equation}
\sin^2 \Theta_W (M_Z) = 0.23126
\label{Theta}
\end{equation}
and under the SM renormalization group equations 
$\sin^2 \Theta_W (\mu)$ increases only slowly with increasing
the scale $\mu$ such that the special value
\begin{equation}
\sin^2 \Theta_W (\mu)  ~~ \equiv ~~ \left( \frac{1}{4} \right)
\label{special}
\end{equation}
is achieved for $\mu \simeq 4$ GeV. To avoid imaginary couplings,
this provides an upper limit for the $331 \rightarrow 321$ symmetry
breaking scale.

\bigskip

\section{The new physics particles}

\bigskip

\noindent
The 331 model \cite{PHF,PP} predicts far fewer new physics particles than
does the competing supersymmetric model, the MSSM. The additional 
331 particles we focus on
here, because they are the simplest to recognize experimentally, are the extra
331 gauge bosons. These are five in number, comprising a neutral $Z^{'}$-boson
and four bileptons\footnote{In 1992 these gauge bosons were named dileptons
\cite{PHF} despite a different usage of the same word by
experimentalists; in 1996 the name was therefore changed to bileptons\cite{PHF3}.}.
They are similar to the $Z^{'}$-bosons in many other models beyond the SM so 
it does not itself provide the most useful 331 smoking gun. 

\bigskip

\noindent
The 331 bileptons, especially those with double electric charge $Y^{++}$
and $Y^{--}$, do provide such a 331 smoking gun. After studying many
different results of proton-proton collisions at $\sqrt{s} = 14$ TeV, within
the context of the 331 model, including the production and decay of the
bileptons as well as the exotic quarks with electric charges $ + \frac{5}{3}$
and $ - \frac {4}{3}$, we wish to point out, as a candidate for the most striking 
331 smoking gun, the existence of X-events.

\bigskip

\noindent
What are X-events? To illustrate the unusual kinematics let us assume the
bilepton mass is $M_Y = 1.5$ TeV. The bileptons are siblings of the familiar
$W^{\pm}$ of the weak interactions and just as for $W^{\pm}$ it is reasonable
to assume a gauge boson mass somewhat below the corresponding spontaneous
symmetry breaking scale.

\bigskip

\noindent
There are also important lower bounds on the bilepton mass. There exist two
such bounds, both arising from table-top experiments and both by coincidence
having been exquisitely measured at the Paul Scherrer Institute (PSI).

\bigskip

\noindent
The first lower mass limit arises from muonium-antimuonium conversion where the
exchange of the doubly-charged bilepton can give the requisite $\Delta L = \pm 2$
violation of lepton number. The experimental result \cite{muoniumantimuonium} provides
a lower limit depending on the 331 gauge coupling $g_{3l}$ and corresponds to
$M_Y > 850$ GeV.

\bigskip

\noindent
A second lower bound on $M_Y$ originates from high precision measurements of
muon decay. The point is that bilepton exchange contributes a $(V+A)$ component
to the familiar $(V-A)$ structure for muon decay predicted by the SM.
This $(V+A)$ component is best measured by the Michel parameter 
and measurements existing for muon decay provide \cite{MichelParameter} a lower
mass limit $M_Y > 1$ TeV, slightly stronger than the lower limit from muonium conversion.

\bigskip

\noindent
Now let us consider proton-proton collisions at 14 TeV and, in particular, the
inclusive process $p+p \rightarrow Y^{++} + Y^{--} + $ anything, with $M(Y^{\pm\pm}) \simeq 1.5$ 
TeV. The kinematics
strongly favor production of the two on-shell bileptons extremely close to at rest
in the center-of mass frame. They then both decay into back-to-back muons,
which explains the name X-event to characterize its appearance.

\bigskip

\noindent
The four muons share a huge Lorentz factor $\gamma = (0.75\times 10^6)/106
\simeq 7,000$ corresponding to a velocity divided by the speed of light equal to
one up to a part in $10^{8}$.  This ultrarelativistic speed raises a practical difficulty 
in detecting the track curvature 
due to the magnetic field in either the ATLAS or the CMS detectors. If this curvature 
could be 
measured, however, it would provide a useful check because the same-sign 
nature of the bilepton decays could be confirmed.

\section{Discussion}

\noindent
The X-event we have discussed in this letter provides a simple smoking gun for the 331 model. 
It is the simplest example of which we are aware. There
are, of course, many other characteristic 331 signatures, some of which 
were discussed in \cite{PHF2}.

\section*{Acknowledgement}

\noindent
We thank Professors G. Leontaris, J. Rizos and K. Tamvakis for organizing the
STRINGPHENO2016 conference in Ioannina which provided a stimulating environment.
One of us (P.H.F.) thanks INFN for hospitality and support at the University of
Salento.

\end{document}